\documentclass[12pt]{article}
\begin{document}

\title{The Analog of t'Hooft Pions with Adjoint Fermions\footnote{This work
is done in collaboration with F. Antonuccio, Max Planck Institiute
Heidelberg, Germany} \footnote{Invited talkat ``New Non-Perturbative
Methods and Quantisation on the Light-Cone', Les Houches, France 24
Feb.-7 March 1997.(to appear in the proceedings} }
\author{Stephen S. Pinsky 
\\ Department of Physics, 
\\ The Ohio State University, 
\\ Columbus, OH 43210}
\maketitle

\section{Introduction}
In this paper we discuss and solve a class of $1+1$
dimensional matrix field theories in the light-cone Hamiltonian
approach that are obtained from a  dimensional reduction of $3+1$
dimensional Yang-Mills theory. Recently a similar procedure has been
used to formulate a conjectured for M theory  \cite{baf96}.

The strategy in formulating the model is to
retain all of the essential degrees of freedom of higher
dimensional QCD. We start by considering ${\mbox{QCD}}_{3+1}$  coupled to
Dirac adjoint fermions\cite{anp97}. The virtual creation of
fermion-antifermion pairs is not suppressed in the large-$N$ limit -- in
contrast to the case for fermions in the fundamental representation
\cite{and96b,and96a,buv95} -- and so one may study the structure
of boundstates beyond the valence quark (or quenched)
approximation.

The gauge group of the theory is actually $SU(N)/Z_N$, which has
nontrivial topology and vacuum structure. For the particular gauge
group $SU(2)$ this has been discussed elsewhere \cite{pir96}. While
this vacuum structure may in fact be relevant for a discussion of
condensates, for the purposes of this calculation it will be
ignored.

In the first section we formulate the $3+1$ dimensional $SU(N)$
Yang-Mills theory and then perform dimensional reduction to obtain
a $1+1$ dimensional matrix field theory. The light-cone
Hamiltonian is then derived for the light-cone gauge $A_- = 0$
following a discussion of the physical degrees of freedom of the
theory. We then discuss the exact massless solutions of the
boundstate integral equations. These massless states have constant
wavefunctions in momentum space and are therefore fundamental excitations
of the the theory.

\section{Definitions}

We first consider $3+1$ dimensional $SU(N)$ Yang-Mills coupled to
a Dirac spinor field whose components transform in the adjoint
representation of $SU(N)$:
\begin{equation}
{\cal L} = \mbox{Tr} \left[ -\frac{1}{4} F_{\mu \nu} F^{\mu \nu}
 + \frac{{\rm i}}{2}
(\bar\Psi\gamma ^{\mu} \buildrel \leftrightarrow \over
D_{\mu}\Psi) - m\bar\Psi \Psi \right] \; ,
\label{3+1theory}
\end{equation}
where $D_{\mu} = \partial_{\mu} + {\rm i}g [ A_\mu,\ \ ] $ and $F_{\mu \nu}
= \partial_{\mu}A_{\nu} - \partial_{\nu} A_{\mu} + {\rm i}g [A_{\mu},
A_{\nu} ]$. We also write $A_{\mu} = A_{\mu}^a \tau ^a$ where $\tau^a$ is
normalized such
that $\mbox{Tr} (\tau^a \tau^b ) = \delta_{ab}$.
The projection operators\footnote{
We use the conventions $\gamma^{\pm} = (\gamma^0 \pm \gamma^3)/
\sqrt{2}$, and $x^{\pm}=(x^0 \pm x^3)/\sqrt{2}$.}
$\Lambda_L,\Lambda_R$ permit a decomposition
of the spinor field $\Psi = \Psi_L +\Psi_R$, where
\begin{equation}
\Lambda_L = {1 \over 2} \gamma^+ \gamma^- ,\quad \Lambda_R = {1 \over 2}
\gamma^- \gamma^+ \quad \mbox{and} \quad
\Psi_L = \Lambda_L \Psi, \quad \Psi_R = \Lambda_R \Psi .
\end{equation}
Inverting the equation of motion for $\Psi_L$, we find
\begin{equation}
\Psi_L ={1 \over 2{\rm i}D_-} \left[ {\rm i}
\gamma^i D_i +m \right ] \gamma^+ \Psi_R
\label{eqnmotion}
\end{equation}
where $i=1,2$ runs over transverse space. Therefore $\Psi_L $ is not an
independent degree of freedom.

Dimensional reduction of the $3+1$ dimensional Lagrangian
(\ref{3+1theory}) is performed by assuming (at the classical level)
that all fields are independent of the transverse coordinates
$x^{\perp}=(x^1,x^2)$: $\partial_{\perp} A_{\mu} =0$ and
$\partial_{\perp} \Psi = 0$.
In the resulting $1+1$ dimensional field theory, the transverse components
$A_{\perp} = (A_1,A_2)$ of the gluon field will be represented
by the $N \times N$ complex matrix fields $\phi_{\pm}$:
\begin{equation}
     \phi_{\pm} = \frac{A_1 \mp {\rm i} A_2}{\sqrt{2}}.
\end{equation}
Here,  $\phi_-$ is just the Hermitian conjugate of
$\phi_+$. When the theory is quantized, $\phi_{\pm}$ will correspond to
$\pm 1$ helicity bosons (respectively).

The components of the Dirac spinor $\Psi$ are the $N \times N$
{\em complex} matrices $u_{\pm}$ and $v_{\pm}$, which
are related to the left and right-moving spinor fields according
to
\begin{equation}
\Psi_R ={1 \over 2^{{1 \over 4}}}
\left ( \begin{array}{c} u_+\\ 0 \\ 0 \\u_- \end{array}
\right )
\quad
\Psi_L ={1 \over 2^{{1 \over 4}}}
\left ( \begin{array}{c} 0 \\ v_+ \\ v_- \\0 \end{array}
\right )
\end{equation}
Adopting the light-cone gauge $A_- = 0$ allows one to explicitly
rewrite the left-moving fermion fields $v_{\pm}$ in terms
of the right-moving fields $u_{\pm}$ and boson fields $\phi_{\pm}$,
by virtue of equation (\ref{eqnmotion}). We may therefore eliminate
$v_{\pm}$ dependence from the field theory. Moreover,
Gauss' Law
\begin{equation}
\partial_-^2 A_+ = g \left( {\rm i}[\phi_+, \partial_- \phi_-] +
                    {\rm i}[\phi_-, \partial_- \phi_+]
 + \{ u_+,u_+^{\dagger} \} + \{ u_-,u_-^{\dagger} \} \right)
\end{equation}
permits one to remove any explicit dependence on $A_+$,
and so the remaining {\em physical}
degrees of freedom of the field theory are represented by
the helicity $\pm \frac{1}{2}$ fermions $u_{\pm}$, and the
helicity $\pm 1$ bosons $\phi_{\pm}$. There are no ghosts
in the quantization scheme adopted here.
In the light-cone frame
the Poincar\'e generators $P^-$ and $P^+$ for the
reduced $1+1$ dimensional field theory are given by
\begin{equation}
P^+=\int ^\infty _{-\infty} dx^- \mbox{Tr}  \biggl[
2 \partial_- \phi_- \cdot \partial_- \phi_+
+ {{\rm i} \over 2} \sum_h \left ( u^\dagger_h \cdot
\partial_- u_h -\partial_-
u^\dagger_h \cdot u_h \right )
\biggr ]
\end{equation}
\begin{equation}
P^- = \int ^\infty _{-\infty} dx^- \mbox{Tr} \biggl[
 m_b^2 \phi_+ \phi_-
-{ g^2 \over 2}{J}^+ {1\over \partial_-^2 }{J}^+ +
{ t g^2 \over 2} \left [\phi_+ , \phi_- \right ]^2
+\sum_h{F}^\dagger_h {1 \over {\rm i}\partial_-} {F}^\dagger_h
\biggr ]
\label{hamiltonian}
\end{equation}
where the sum $\sum_h$ is over $h=\pm$ helicity labels,
and
\begin{eqnarray}
J^+ & = & {\rm i}[\phi_+, \partial_- \phi_-] +
                    {\rm i}[\phi_-, \partial_- \phi_+]
 + \{ u_+,u_+^{\dagger} \} + \{ u_-,u_-^{\dagger} \} \\
F_{\pm} & = & \mp s g\left[ \phi_{\pm} \; , u_{\mp} \right ] +
{ m \over \sqrt{2}} u_{\pm} \label{Fterm}
\end{eqnarray}
We have generalized
the couplings
by introducing the variables $t$ and $s$, which do not
spoil the $1+1$ dimensional gauge invariance of the reduced theory;
the variable $t$ will determine the strength of the quartic-like interactions,
and the variable $s$ will determine the
strength of the Yukawa interactions between the fermion and boson fields,
and appears explicitly in equation (\ref{Fterm}). The dimensional
reduction of the original $3+1$ dimensional theory yields the canonical
values $s=t=1$.

Renormalizability of the reduced theory also requires
the addition of a
bare coupling $m_b$, which leaves
the $1+1$ dimensional gauge invariance intact.
In all calculations, the renormalized
boson mass ${\tilde m}_b$ will be set to zero.

Canonical quantization of the field theory
is performed by decomposing the boson
and fermion fields into Fourier expansions
at fixed light-cone time $x^+ = 0$:
\begin{equation}
u_\pm  = {1 \over  \sqrt {2\pi}} \int_{-\infty}^\infty dk
\hspace{1mm} b_\pm(k)  e^{-{\rm i}k x^-} \quad \mbox{and} \quad
\phi_{\pm} = {1 \over  \sqrt {2\pi}} \int_{-\infty}^\infty { dk \over
\sqrt{2|k|}}\hspace{1mm}a_{\pm}(k) e^{-{\rm i}k x^-}
\end{equation}
where $b_{\pm} = b_{\pm}^a\tau^a$ etc.
We also define
\begin{equation}
\quad b_{\pm}(-k) =d_{\mp}^{\dagger}(k),
\quad a_\pm(-k) = a_{\mp}^\dagger (k),
 \label{note}
\end{equation}
where $d_{\pm}$
correspond to antifermions.
Note that
$(b^{\dagger}_{\pm})_{ij}$
should be distinguished
from $b_{\pm ij}^{\dagger}$, since in the former the quantum conjugate
operator $\dagger$ acts on (color) indices, while
it does not in the latter. The latter formalism
is sometimes customary in the study of matrix models.
The precise connection between the usual gauge theory and matrix theory
formalism may be stated as follows:
\[
 b_{\pm ji}^{\dagger} =
b_{\pm}^{a\dagger}\tau^{a*}_{ji}=b_{\pm}^{a\dagger}\tau^a_{ij} =
(b_{\pm}^{\dagger})_{ij}
\]
The commutation and anti-commutation relations (in matrix formalism)
for the boson and fermion fields take the following
form in the large-$N$ limit ($k,{\tilde k} >0$; $h,h'={\pm}$):
\begin{equation}
\left [ a_{h ij}(k),a_{h' kl}^{\dagger}({\tilde k}) \right] =
\{ b_{h ij}(k), b_{h' kl}^{\dagger}({\tilde k}) \}
= \{ d_{h ij}(k), d_{h' kl}^{\dagger}({\tilde k}) \}
= \delta _{h h'} \delta_{jl}\delta_{ik} \delta(k-{\tilde k}),
\label{rhccrs}
\end{equation}
where have used the relation
$\tau^a_{ij} \tau^a_{kl} =
\delta_{il} \delta_{jk} -{1 \over N} \delta_{ij} \delta_{kl}$.
All other (anti)commutators vanish.

The Fock space of physical states
is generated by the color singlet states, which have a natural
`closed-string' interpretation. They are formed by
a color trace  of the fermion, antifermion and boson
operators
acting on the vacuum state $|0\rangle$.
Multiple string states couple to the theory with
strength $1/N$,
and so may be ignored.

\section{The Light-Cone Hamiltonian}
The current-current term $J^+
\frac{1}{\partial_-^2} J^+$ in equation (\ref{hamiltonian})
in momentum space, for example, takes the form
\begin{eqnarray}
P^-_{J^+ \cdot J^+} &=& {g^2 \over 2 \pi}
\int^\infty_{-\infty} dk_1 dk_2 dk_3 dk_4 {\delta(k_1+k_2-k_3-k_4)
\over (k_3-k_1)^2}
{\mbox{Tr} \over 2} \biggl[\nonumber \\
 & & \sum_{h,h'}
:\{b^\dagger_h(k_1),b_h(k_3)\}:
:\{b^\dagger_{h'}(k_2),b_{h'}(k_4)\}: \nonumber \\
&+ &{(k_1+k_3)(k_2+k_4) \over 4 \sqrt{|k_1||k_2||k_3||k_4|}}
:[a_+^\dagger(k_1),a_+(k_3)]:
:[a_{+}^\dagger(k_2),a_{+}(k_4)]: \nonumber \\
&+ &{(k_2+k_4) \over 2 \sqrt{|k_2||k_4|}} \sum_{h}
:\{ b^\dagger_h(k_1),b_h(k_3)\}:
:[a_{+}^\dagger(k_2),a_{+}(k_4)]: \nonumber \\
&+ &{(k_3+k_1) \over 2\sqrt{|k_1||k_3|}} \sum_{h'}
:[ a_+^\dagger(k_1),a_+(k_3)]::\{b^\dagger_{h'}(k_2),b_{h'}(k_4)\}:\biggr]
\label{jj}
\end{eqnarray}
The explicit form of remaining terms of the Hamiltonian
(\ref{jj}) in terms of the operators $b_{\pm}$,
$d_{\pm}$ and $a_{\pm}$ is straightforward to calculate, but too long
to be written down here. It should be stressed, however,
that several $2 \rightarrow 2$ parton
processes are suppressed by a factor $1/N$, and so are ignored in
the large-$N$ limit.
No terms involving $1 \leftrightarrow 3$ parton interactions are
suppressed in this limit, however.

One  can show that this Hamiltonian  conserves total helicity $h$, which
is an additive quantum number. Moreover, the number of fermions  {\em
minus} the number of antifermions is also conserved in each interaction,
and so we have an additional quantum number ${\cal N}$.  States with
${\cal N} = even$  will be referred to as {\em boson} boundstates, while
the quantum number  ${\cal N} = odd$ will refer to {\em fermion}
boundstates. The cases ${\cal
N}= 0$ and $3$, are analogous to 
conventional mesons
and baryons (respectively).

\section{Exact Solutions}
For the special case
 $s=m={\tilde m}_b = 0$, the only
surviving terms in the Hamiltonian
(\ref{hamiltonian}) are the current-current interactions
$J^+ \frac{1}{\partial_-^2} J^+$ and the four gluon interaction $ \left
[\phi_+ , \phi_- \right ]^2$. The current-current interaction includes
four gluon interaction, four Fermion interaction and two gluon, two Fermion
interactions. This theory  has infinitely many massless boundstates, and the
partons in these states are either fermions or antifermions.
States with bosonic $a_{\pm}$ quanta are always
massive. One also finds that the massless states are pure,
in the sense that the number of partons is a fixed integer,
and there is no mixing between sectors of different parton number.
In particular, for each integer $n \geq 2$, one can always find a massless
boundstate consisting of a superposition
of only $n$-parton states.  A striking feature
is that the wavefunctions of these states
are {\em constant}, and so these states are natural
generalizations of the constant wavefunction solution appearing in
t'Hooft's model \cite{tho74}.

We present an explicit examples below of such a constant
wavefunction solution involving a three fermion state with
total helicity $+\frac{3}{2}$, which is perhaps the simplest
case to study. We apply $P^-$ to the state and show that the result vanishes.
Massless states with five or more partons
appear to have more than one wavefunction which are non-zero
and constant, and in general the wavefunctions are unequal.
It would be interesting to classify all states systematically, and
we leave this to future work. One can, however, easily
count the number of massless states. In particular,
 for ${\cal N }=3$, $h=+\frac{3}{2}$ states, there is
one three-parton state, $2$ five-parton
states, $14$ seven-parton
states and $106$ nine-parton states that yield massless solutions.

Let us now consider the action of the light-cone
Hamiltonian $P^-$ on the three-parton state
\begin{eqnarray}
&&|b_+ b_+ b_+ \rangle =
\int_0^{\infty} dk_1 dk_2 dk_3
\hspace{1mm} \delta (\sum_{i=1}^3 k_i - P^+) \nonumber \\
 &&f_{b_+ b_+ b_+}(k_1,k_2,k_3)
\frac{1}{N^{3/2}} \mbox{Tr}[b_+^{\dagger}(k_1)
 b_+^{\dagger}(k_2)b_+^{\dagger}(k_3)]|0\rangle
\end{eqnarray}
The quantum number ${\cal N}$ is 3 in this case, and ensures
that the state $P^-|b_+ b_+ b_+ \rangle$ must have at least three
partons. In fact, one can deduce the following:
\begin{eqnarray}
& & P^- \mid b_+ b_+ b_+\rangle =
\int_0^{\infty} dk_1 dk_2 dk_3
\hspace{1mm} \delta (\sum_{i=1}^3 k_i - P^+) \times \nonumber \\
\lefteqn{\left\{ -{g^2 N \over 2\pi} \right. \int^\infty_0 d\alpha d\beta
\frac{\delta (\alpha + \beta - k_1 -k_2)}
{(\alpha - k_1)^2} \left[ f_{b_+ b_+ b_+ }(\alpha,\beta,k_3)
- f_{b_+ b_+ b_+}(k_1,k_2,k_3) \right] } & & \nonumber \\
& & \frac{1}{N^{3/2}}\mbox{Tr}
\left[b_+^\dagger(\alpha)b_+^\dagger(\beta)b_+^\dagger(k_3)\right]
\mid 0 \rangle  \nonumber \\
& &+{{g^2 N}\over 2\pi}\int^\infty_{0}
d\alpha d\beta d\gamma \sum_h
{\delta(\alpha + \beta + \gamma - k_1)
\over (\alpha+\beta)^2}
f_{b_+ b_+ b_+}(\alpha + \beta + \gamma ,k_2,k_3) \frac{1}{N^{5/2}} 
\nonumber \\
\lefteqn{\mbox{Tr}\left[
\{b_h^\dagger(\alpha),d_{-h}^\dagger(\beta)\}b_+^\dagger(\gamma)
b_+^\dagger(k_2)b_+^\dagger(k_3)-
 \{b_h^\dagger(\alpha),d_{-h}^\dagger(\beta)\}
b_+^\dagger(k_2)b_+^\dagger(k_3)b_+^\dagger(\gamma)
\right] \mid 0 \rangle} & & \nonumber \\
& &\nonumber \\
& &+{{g^2 N} \over 4\pi}\int^\infty_{0} d\alpha d\beta d\gamma
  \sum_h
{\delta(\alpha + \beta + \gamma - k_1 )
\over\sqrt{\alpha \beta}(\alpha+\beta)^2 }f_{b_+ b_+ b_+}\frac{1}{N^{5/2}}
(\alpha + \beta + \gamma, k_2,k_3)
 \nonumber \\
\lefteqn{\mbox{Tr}\left[
[a_h^\dagger(\alpha),a_{-h}^\dagger(\beta)]b_+^\dagger(\gamma)
b_+^\dagger(k_2)b_+^\dagger(k_3)-
 [a_h^\dagger(\alpha),a_{-h}^\dagger(\beta)]
b_+^\dagger(k_2)b_+^\dagger(k_3)b_+^\dagger(\gamma)
\right] \mid 0 \rangle} & & \nonumber \\
& + & \mbox{ cyclic permutations} \left. \frac{}{} \right\}
\label{exacts}
\end{eqnarray}
The five-parton states  above
correspond to virtual fermion-antifermion and boson-boson
pair creation.
The expression (\ref{exacts}) vanishes if the wavefunction
$f_{b_+ b_+ b_+}$ is constant.

One of the exact seven particle wave functions is
\begin{eqnarray}
&&\int_0^{\infty} dk_1 dk_2 dk_3 dk_4 dk_5 dk_6 dk_7
\hspace{1mm} \delta (\sum_{i=1}^7 k_i - P^+)
 \frac{1}{N^{7/2}} { 1 \over \sqrt{2}}\nonumber \\ 
&& \mbox{Tr}[ 
b_+^{\dagger}(k_1)b_+^{\dagger}(k_2) b_+^{\dagger}(k_3)
b_-^{\dagger}(k_4) b_-^{\dagger}(k_5) d_+^{\dagger}(k_6)
d_+^{\dagger}(k_7)-\\ &&
b_+^{\dagger}(k_1)b_+^{\dagger}(k_2) b_+{\dagger}(k_3)  
b_-^{\dagger}(k_4)d_-^{\dagger}(k_5) b_+^{\dagger}(k_6)d_+^{\dagger}(k_7)]
|0\rangle \nonumber
\end{eqnarray}

\section{ Conclusions }

We have presented a non-perturbative Hamiltonian formulation of a class of
$1+1$ dimensional matrix field theories, which may be derived from a
classical dimensional reduction of
${\mbox{QCD}}_{3+1}$ coupled to Dirac adjoint fermions. We choose to adopt
the light-cone gauge $A_- = 0$, and solve the boundstate integral equations
in the large-$N$ limit. Different states may be classified according to total
helicity $h$, and the quantum number ${\cal N}$, which defines the number of
fermions minus the number of antifermions in a state.

For a special choice of couplings  we find an infinite number of pure
massless states of arbitrary length. The wavefunctions of these states are
always constant, and are the analogs of the t' Hooft pion. Sometimes the
wavefunctions  involves several (possibly different) constants. In this model
we have both fermion and boson states of this type. The state explicitly
shown here are all fermions. 

The techniques employed here are not specific to the choice of field theory,
and are expected to have a wide range of applicability, particularly in the
light-cone Hamiltonian formulation of supersymmetric field
theories\cite{maa95,hak95}.

\section{acknowledgments}
\noindent The work was supported in part by a grant from
the US Department of Energy. Travel support was provided in part by a NATO
collaborative grant.

\end{document}